\documentclass[aps,prl,groupedaddress,twocolumn,floatfix,showpacs,tightenlines]{revtex4-1}
\usepackage{graphicx}
\usepackage{color}

\newcommand{\be}{\begin{equation}}
\newcommand{\ee}{\end{equation}}
\newcommand{\bea}{\begin{eqnarray}}
\newcommand{\eea}{\end{eqnarray}}
\newcommand{\bes}{\begin{subequations}}
\newcommand{\ees}{\end{subequations}}

\begin{document}
\title{Flip-flopping binary black holes}
\author{Carlos O. Lousto}
\author{James Healy} 
\affiliation{Center for Computational Relativity and Gravitation,\\
School of Mathematical Sciences,
Rochester Institute of Technology, 85 Lomb Memorial Drive, Rochester,
 New York 14623}
\date{\today}

\begin{abstract}
We study binary spinning black holes to display the long term individual
spin dynamics. We perform a full numerical simulation starting
at an initial proper separation of $d\approx25M$ between equal mass holes
and evolve them down to merger for nearly 48 orbits,
3 precession cycles, and half of a flip-flop cycle. The simulation lasts for
$t=20000M$ and displays a total change in the orientation of the spin of
one of the black holes from initially aligned with the orbital 
angular momentum to a complete anti-alignment after half of a flip-flop
cycle. We compare this evolution with an integration of the 3.5 
Post-Newtonian equations of motion and spin evolution to show that 
this process continuously flip-flops the spin during the lifetime of 
the binary until merger. We also provide
lower order analytic expressions for the maximum flip-flop angle and
frequency. We discuss the effects this dynamics may have on spin growth in 
accreting binaries and on the observational consequences for galactic and 
supermassive binary black holes.
\end{abstract}

\maketitle

{\it Introduction:}
Numerical relativity techniques are now able
to directly simulate binary black hole  mergers 
\cite{Pretorius:2005gq, Campanelli:2005dd, Baker:2005vv}. 
In particular one can follow the dynamics
of black hole spins in an inspiral
orbit down to the formation of the
final remnant black hole \cite{Campanelli:2006uy}. 
One of the most striking results of those studies
has been the discovery of
very large recoil velocities \cite{Campanelli:2007ew}
acquired by the merger remnant, up to $5000km/s$ \cite{Lousto:2011kp}
for hangup configurations.

It has been pointed out \cite{Bogdanovic:2007hp} 
that the presence of accreting
matter can align (or counter align) spins with the orbital
angular momentum thus reducing the recoil velocities to a few
hundred $km/s$ ~\cite{Healy:2014yta}. 
Recent studies of the tidal effects on tilted 
accretion disks around spinning black holes find almost
perfect alignments of the spins with the orbital angular momentum
\cite{Miller:2013gya,Sorathia:2013pca} 
on a shorter time
scale than that of gravitational radiation (for black hole separations
above a thousand gravitational radii).

While those studies have been performed on individual black holes,
we revisit this scenario to study the precession dynamics of black hole
spins in a binary system. In particular we are interested in
the dynamics of polar precession of each individual
spin. We find a flip-flop mode with periods shorter than the
gravitational radiation scale and with relatively high probability
to occur given generic (but comparable mass) initial configurations. 
We briefly 
discuss the effects that this flip-flopping spin could have on the inner
accretion disk dynamics and its potential observational consequences.

{\it Full Numerical Evolution:}
In order to display the long term dynamics of spinning binary black
holes in General Relativity we start a numerical simulation at a proper 
separation $d\approx25M$.
We study an equal mass binary with 
different spin magnitudes and orientations. In particular, we
choose one of the black holes as slowly spinning with its spin $\vec{S}_1$ 
initially aligned with the orbital angular momentum $\vec{L}$, while
the second highly spinning black hole has spin $\vec{S}_2$ 
lying mostly along the orbital plane, but slightly anti-aligned with 
$\vec{L}$, such that the total spin $\vec{S}$ exactly lies in the orbital 
plane, i.e. $\vec{S}\cdot\vec{L}=0$.
These choices (See Table~ \ref{tab:ID})
are for the sake of simplicity of the analysis, and also
provide a plausible scenario where accretion has proceeded to align one of the
black holes with $\vec{L}$ and led to comparable masses by preferably
accreting onto the initially smaller hole \cite{Farris:2013uqa}. 
We also choose the magnitude
of the first black hole to be smaller than that of the second, foreseeing
(as discussed later in this paper)
that the flip-flopping spin neutralizes (at least partially) the
growth of intrinsic spin magnitudes, $S_{1,2}/m_{1,2}^2$, by accretion.

\begin{table}
\caption{Initial data parameters and system details.  The punctures are located
at $\vec r_1 = (x_1,0,z)$ and $\vec r_2 = (x_2,0,z)$, with momenta
$P=\pm (0, P,0)$, spins $\vec S_1 = (0, 0, S_{1z})$ and $\vec S_2 = (S_{2x}, 0, S_{2z} )$, 
mass parameters $m^p$, horizon (Christodoulou) masses $m^H$, total ADM mass
$M_{\rm ADM}$, and dimensionless spins $\alpha = a/m_H = S/m_H^2$. The horizon
masses and spins are given after the gauge settles, and the
errors in $m^H$ and $\alpha$ are determined by the drift in the quantity 
during the inspiral.  Also provided are the simple proper distance $d$,
eccentricity at the start of the inspiral $e_i$, 
and eccentricity $e_f$ and the number of orbits $N$ just before merger.
}
\label{tab:ID}
\begin{ruledtabular}
\begin{tabular}{ccccc}
$x_1/m$ & $x_2/m$  & $z/m$ & $P/m$ & $d/m$\\
10.73983 & -10.76016 & -0.01968 & 0.05909 & 25.37 \\
\hline
$m^p_1/m$ & $m^p_2/m$ & $S_{1z}/m^2$ & $S_{2x}/m^2$ & $S_{2z}/m^2$\\
0.48543 & 0.30697 & 0.05 & 0.19365 & -0.05 \\
\hline
$M_{\rm ADM}/m$ & $J_{\rm ADM}/m^2$ & $e_i$ & $e_f$ & $N$ \\
0.99472 & 1.2704344 & 0.0322 & 0.0006 & 48.5 \\
\hline
$m^H_1/m$ & $\delta m^H_1/m$ & $m^H_2/m$ & $\delta m^H_2/m$ & \\
0.50000 & 0.00002 & 0.49974 & 0.00001 & \\
\hline
$\alpha_1$ & $\delta \alpha_1 $ & $\alpha_2$ & $\delta \alpha_2$& \\
0.20003 & 0.00056 & 0.80088 & 0.00066 & \\
\end{tabular}
\end{ruledtabular}
\end{table}

We use the TwoPunctures thorn~\cite{Ansorg:2004ds}, a spectral
numerical code to generate initial
``puncture'' (no excision of the horizon) data for the binary black hole
simulations. We evolve these initial data sets 
using the {\sc LazEv}~\cite{Zlochower:2005bj} implementation of the 
moving puncture approach~\cite{Campanelli:2005dd}.
For the runs presented here, we use centered, eighth-order finite 
differencing in space~\cite{Lousto:2007rj} and a fourth-order Runge Kutta time
integrator. Our code
uses the {\sc Cactus}/{\sc EinsteinToolkit}~\cite{cactus_web,
Loffler:2011ay} infrastructure for parallelization.  We use the {\sc
Carpet}~\cite{Schnetter-etal-03b} mesh refinement driver to provide a
``moving boxes'' style of mesh refinement.
We locate the apparent horizons using the {\sc AHFinderDirect}
code~\cite{Thornburg2003:AH-finding} and measure the horizon spin
using the isolated horizon algorithm detailed
in~\cite{Dreyer02a}. For the computation of the radiated energy 
and linear momentum
we use the asymptotic formulas in \cite{Campanelli:1998jv} which are expressed
directly in terms of the Weyl scalar $\psi_4$.

To complete the full evolution required $2.5$ million service units on 25 
to 30 nodes of our local cluster ``Blue Sky'' with dual Intel Xeon E5-2680 
processors nearing $100M$ of evolution per day. 
Our evolution is free and we verify its accuracy by the satisfaction of
the Hamiltonian and Momentum constraints. All four $L_2$-norm quantities 
remain well below $10^{-8}$ until merger.
Individual horizon masses $m^H_1$ and $m^H_2$
are preserved to a level of 2 and 1.4 parts in $10^{5}$ respectively
until merger. Spins grow linearly with time until merger by a total
increase of $1.5\times10^{-4}$. Thus the total increase of the
intrinsic spin magnitudes $\alpha_{1,2}=S_{1,2}/m_{1,2}^2$ are 
$\delta\alpha_1=6\times10^{-4}$
and $\delta\alpha_2=6\times10^{-4}$ from initial data to merger,
as described in Table~\ref{tab:ID}.

\begin{figure}
\centerline{
\includegraphics[width=0.49\columnwidth]{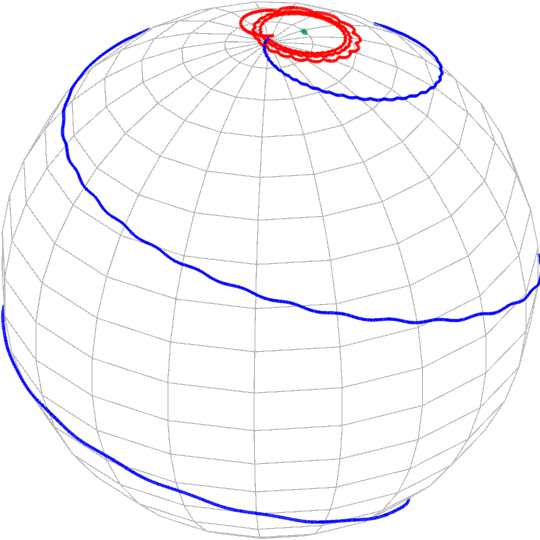}
\includegraphics[width=0.49\columnwidth]{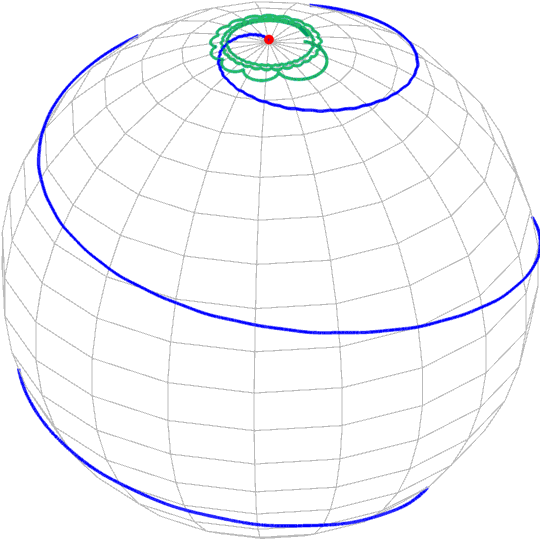}}
\caption{Directional evolutions of the spins and angular momentum in
the initial coordinate frame (left) and in the noninertial $\vec{L}$
frame (right). Color Keys: red $\hat{L}$,
green $\hat{J}$, blue $\hat{S}_1$.
\label{spheres}}
\end{figure}

The azimuthal precessional effect and polar flip-flop can be directly
seen in the evolution of the spin components of the black holes represented
over a sphere in Fig.~\ref{spheres}. The effect is apparent
in the frame of the orbital plane as well as the fixed initial set of 
coordinates.

\begin{figure}
\includegraphics[angle=0,width=\columnwidth]{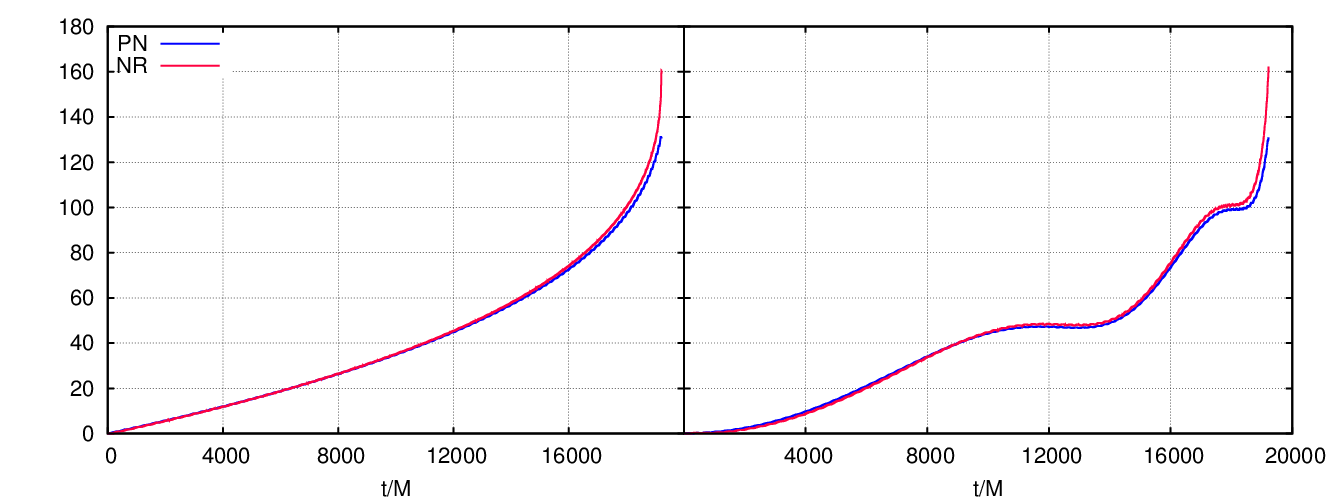}
\caption{The angle between the spin of the secondary (smaller spin) black hole
$\vec{S}_1$ with respect to the orbital angular momentum $\vec{L}$ (left)
and with respect to the fixed $z$-axis (right). For comparison we also plot
the 3.5PN prediction, which underestimates the flip of the angle at the 
latest stages of evolution (merger).
\label{fig:S1anges}}
\end{figure}

Fig.~\ref{fig:S1anges} displays the angles that the 
(slower spinning) black hole
spin $\vec{S}_1$ forms with the precessing orbital 
angular momentum $\vec{L}$ or with the fixed $\hat{z}$-axis
as a function of time. Both start originally aligned and by the time
of merger both display an almost total flip, around $160^\circ$. 
Had we started the binary
further separated apart this spin would continue to flip-flop between complete
alignment and counter alignment as described in the next section using the
Post-Newtonian (PN) approximation.
We also compare our results with the corresponding 
3.5PN integration of the equations
of motion and spin evolution~\cite{Damour:2007nc,Buonanno:2005xu}. 
We observe a long initial superposition of the 
PN and full numerical precession curves corresponding to the early 
$15000M$ of evolution, when the
the binary's separations is above around $15M$. As the merger proceeds
and the evolution becomes more dynamical we observe larger
deviations from each other, with the 
numerical solution to General Relativity
presenting a stronger spin-flip effect.

\begin{figure}
\includegraphics[angle=0,width=\columnwidth]{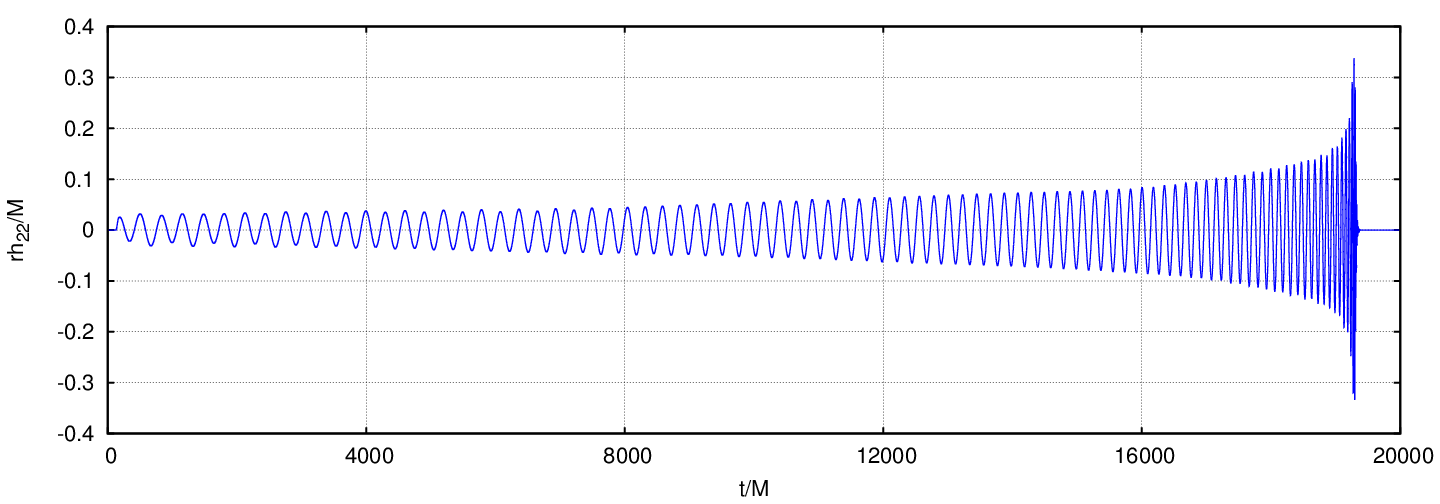}
\includegraphics[angle=0,width=\columnwidth]{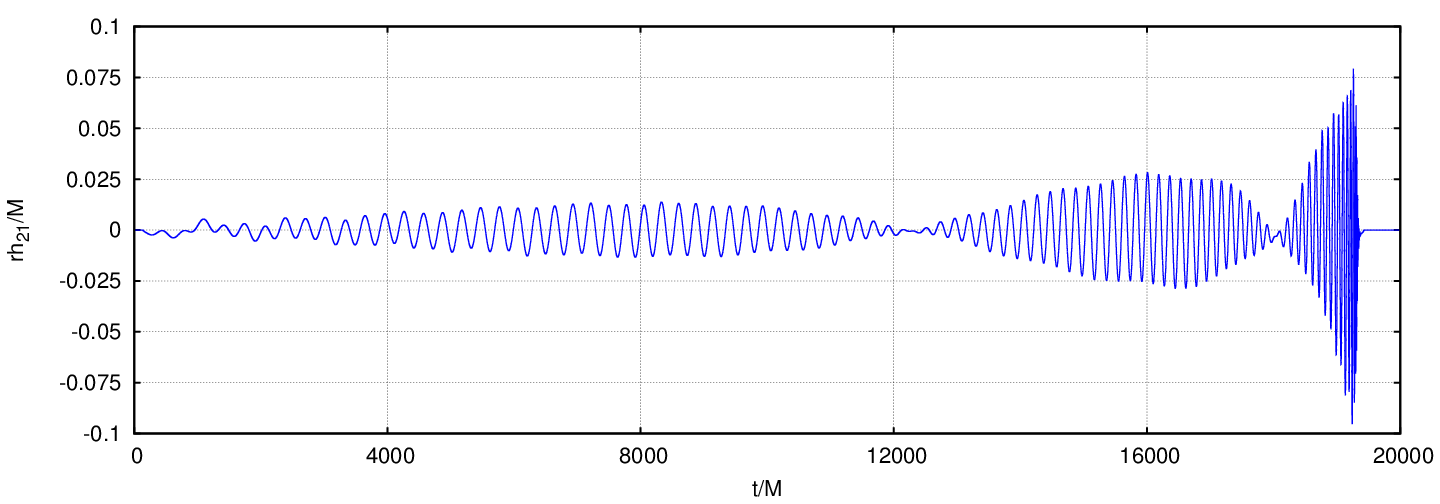}
\caption{The real part of the waveform strain for the modes
$(\ell,m)=(2,2)$ and $(\ell,m)=(2,1)$. While the former (top)
gives the leading chirping 
amplitude, the latter (bottom)
clearly displays the precession effect, completing nearly
three cycles during the $t=20000M$ of the simulation.
\label{fig:h22h21}}
\end{figure}

Fig.~\ref{fig:h22h21} displays the leading waveform modes for 
the strain. In the top panel is the characteristic chirp of the 
$(\ell,m)=(2,2)$ mode, with an increasing amplitude slightly 
modulated at around the orbital frequency due to the nutation of
$\vec{L}$ around the total angular momentum $\vec{J}$ 
(See Fig. 1 in Ref. \cite{Lousto:2013vpa}). The lower panel shows 
the azimuthal precessional effect of 
$\vec{L}$ on the amplitude of the $(2,1)$ mode, showing (in a gauge
invariant way) that we evolved for
nearly three precessional cycles (See Ref.~\cite{Campanelli:2008nk} 
for a first discussion relating this mode to precession in full numerical
simulations).

Table~\ref{tab:remnant} displays the properties of the final
black hole remnant formed after merger. Notably, the recoil reaches 
$1500km/s$, and the orientation of the
final spin changes by only $1.62$ degrees
with respect to the initial direction of the total angular momentum,
as expected for comparable mass binaries~\cite{Barausse:2009uz,Lousto:2013wta}.

\begin{table}
\caption{Remnant properties and recoil velocity.
The final mass and spin are measured from the horizon, and the
recoil velocity is calculated from the gravitational waveforms.
The error in the mass and spin is determined by the drift in 
those quantities after the remnant settles down.  The error in
the recoil velocity is the difference between first and second
order polynomial extrapolation to infinity.
\label{tab:remnant}}
\begin{ruledtabular}
\begin{tabular}{ccc}
$M_{rem}/m$ & $|\alpha_{rem}|$  & $V_{recoil}[km/s]$ \\
$0.94904 \pm 0.00000$ & $0.70377 \pm 0.00002$ & $1508.49 \pm 16.08$ \\
\hline
$\alpha_{rem}^x$ & $\alpha_{rem}^y$  & $\alpha_{rem}^z$ \\
$0.10815\pm0.00003$ & $-0.01986\pm0.00000$ & $0.69513\pm0.00002$ \\
\end{tabular}
\end{ruledtabular}
\end{table}

{\it Post Newtonian spin dynamics:}
In order to provide an analytic understanding of the flip-flop
spin mode, we look at the precession equations for the spins 
$\vec{S}_1$ and $\vec{S}_2$ with a mass ratio $q=m_1/m_2$
to leading spin-orbit and spin-spin couplings in 
the (2PN) post-Newtonian expansion~\cite{Buonanno:2005xu}
\bea\label{spinevo}
\frac{d\vec{S}_1}{dt}&=&
\frac{1}{r^3}\left[\left(2+\frac{3}{2q}\right)\vec{L}-\vec{S}_2
+\frac{3( \vec{S}_0\cdot\hat{n})}{1+q}\hat{n}\right]\times\vec{S}_1,\nonumber\\
\frac{d\vec{S}_2}{dt}&=&
\frac{1}{r^3}\left[\left(2+\frac{3q}{2}\right)\vec{L}-\vec{S}_1
+\frac{3q(\vec{S}_0\cdot\hat{n})}{1+q}\hat{n}\right]\times\vec{S}_2,\,
\eea
where $\vec{n}=\vec{r}_1-\vec{r}_2$ and
\be
\vec{S}_0=\left(1+\frac1q\right)\vec{S}_1+(1+q)\vec{S}_2.
\ee
For more details see the reviews in 
Refs.~\cite{Blanchet:2013haa,Schafer:2014cxa}.

\begin{figure}
\centerline{
\includegraphics[width=1.25in]{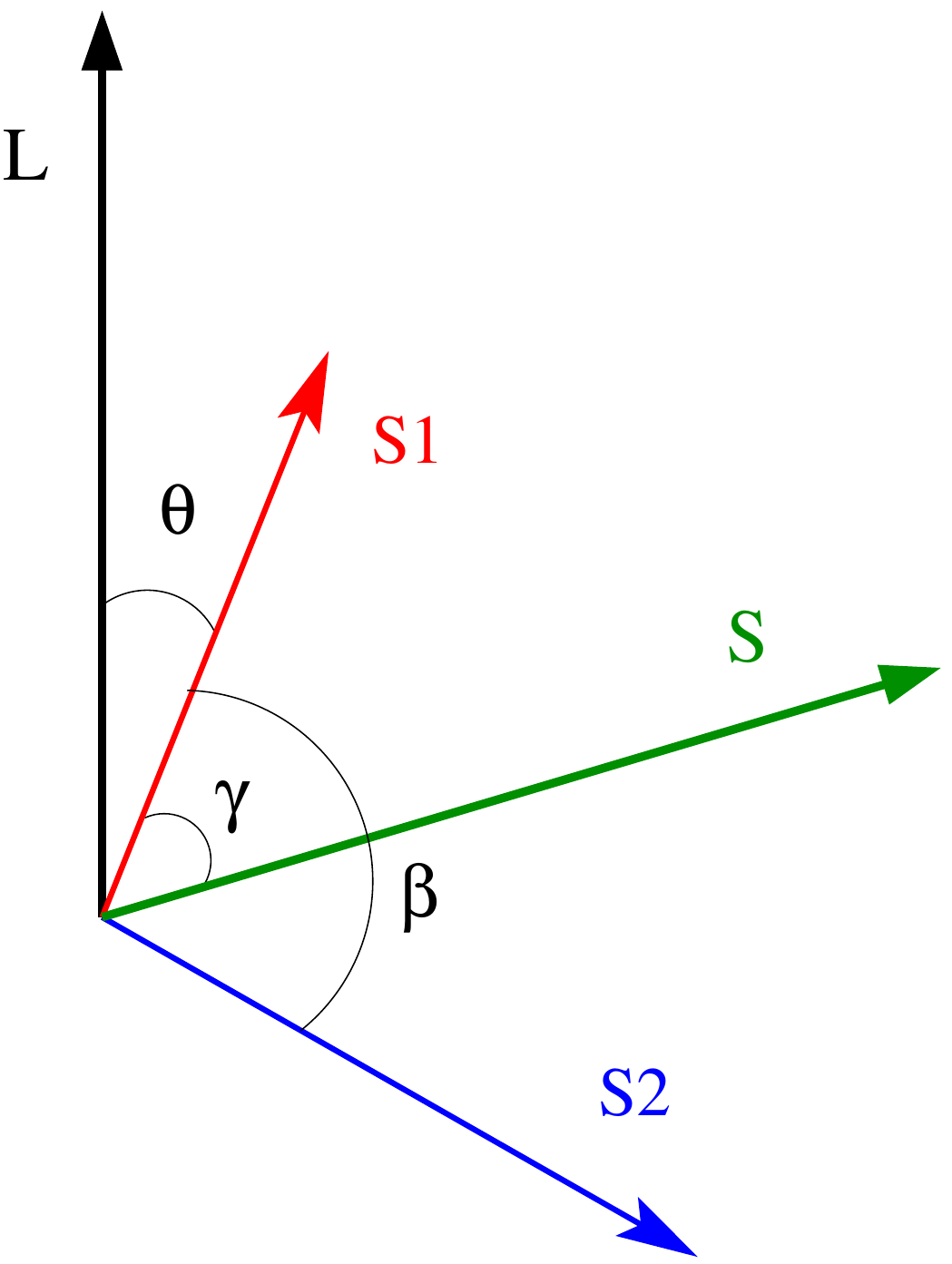}}
\caption{Spin configurations $\vec{S}_1$ and $\vec{S}_2$
relative to the orbital angular momentum $\vec{L}$. 
Here $\vec{S}=\vec{S}_1+\vec{S}_2$.\label{spinconfig}}
\end{figure}

For direct connection with the full numerical simulation above 
we will consider here the equal mass case, i.e. $q=1$ and for the sake
of simplicity, the conservative 2PN spin dynamics at fixed $r$.
We next consider a generic configuration of binary black holes
with arbitrary spins
$\vec{S}_1$ and $\vec{S}_2$ at an angle $\beta$ with respect to
each other and adding up to the vector $\vec{S}$. For definiteness
$\vec{S}_1$ is the spin of the black hole $1$ at an angle
$\gamma$ with respect to $\vec{S}$ as shown in Fig.~\ref{spinconfig}
and $\vec{S}_2$ is the spin of the black hole $2$ identified with the
larger spin magnitude $S_2$.

From Eqs.~(\ref{spinevo}) the magnitude of the individual spins
$S_1$ and $S_2$ are conserved as well as the magnitude of its
sum, $S$ (This has been observed to be approximately true 
in full nonlinear simulations of binary black holes solving
General Relativity field equations numerically~\cite{Campanelli:2006fg}). 
It follows that the following quantities are conserved:
\bea
\vec{S}\cdot\vec{S}&=&S^2=S_1^2+S_2^2+2S_1S_2\cos\beta=\text{constant,}\\
\vec{S}\cdot\vec{S}_1&=&SS_1\cos\gamma=S_1^2+S_2S_1\cos\beta=\text{constant.}
\eea
In turn, this leads to the conservation of $\beta$ and $\gamma$ during the evolution of
the binary. In particular we find that $\vec{S}_1$ oscillates 
around $\vec{S}$ between polar
angles $\gamma$ and $-\gamma$ (when it is both coplanar to 
$\vec{S}$ and $\vec{L}$). We call this the {\it flip-flop} angle
\be
\theta_{ff}=\theta_{max}-\theta_{min}=2\gamma,
\ee
where
\be\label{gamma}
\cos\gamma=\frac{S_1+S_2\cos\beta}{\sqrt{S_1^2+S_2^2+2S_1S_2\cos\beta}}
=\frac{S^2+S_1^2-S_2^2}{2SS_1}.
\ee

By decomposing the spin evolution equations (\ref{spinevo}) along
$\vec{L}$ and perpendicular to it, in the fashion of \cite{Lousto:2013wta},
Sec.~IV.A, we obtain equations of the form
$d(\vec{S}_i\cdot\hat{L})/dt=\Omega_{ff}\,\vec{S}_i\cdot\hat{L}+...$
for $i=1,2$ and analogously for the perpendicular component of $S_i$
giving $\Omega_p$.
From where we can read-off the average polar and azimuthal oscillations 
frequencies of the spin $\vec{S}_1$ (See also \cite{Racine:2008qv})
\bea\label{Omegaff}
\Omega_{ff}&=&3\frac{S}{r^3}\left[1-\frac{2\,\vec{S}\cdot\hat{L}}{M^{3/2}r^{1/2}}\right],\\
\Omega_{p}&=&\frac{7L}{2r^3}+\frac{2}{r^3}(\vec{S}\cdot\hat{L}).
\eea
that we identify with the flip-flop and precession frequencies respectively.

Note that the black hole $2$ also oscillates at this $\Omega_{ff}$ 
frequency, but with a smaller flip-flop angle (Since $S_2>S_1$)
given by $2(\beta-\gamma)$ where
\be
\cos(\beta-\gamma)=\frac{S^2+S_2^2-S_1^2}{2SS_2}.
\ee
Thus both spins, $\vec{S}_1$ and $\vec{S}_2$, oscillate around $\vec{S}$ 
which in turn precess around $\vec{L}$.

This oscillation of the spins represent a {\it genuine} spin-flip in the
sense that it is the same object that completely changes its spin orientation. 
This is different from the simple case where the final remnant spin has flipped
direction when compared to the spin of one of the individual orbiting
black holes \cite{Campanelli:2006fy}.

{\it Discussion:}
In the scenario of binary black holes carrying individual
thin accretion disks (and possibly a common circumbinary disk), 
spins changing their orientation can generate
dramatic dynamical effects on the accreting matter around them. 
For definiteness, we focus on the black hole
with spin $\vec{S}_1$ undergoing direction changes, 
which when viewed in the orbital frame, resembles the peeling
of an orange (See Fig.~\ref{spheres}).
Due to the relatively short time scale of flip-flop 
at close separations, the accreting matter increases the black
hole spin during half the flip-flop period, but decreases it during
the other half. On the other hand, mass is always added to the
black holes during both the up and down states.  The resulting net 
effect is to lower the intrinsic spin magnitude, $S_1/m_1^2$.

From Eqs.~(\ref{gamma}) and (\ref{Omegaff}),
which represent a good approximation for well separated binaries $(r>>100M)$,
requiring a flip-flop angle of $180^\circ$ implies that $\gamma=\pi/2$ and
$\Omega_{ff}=3\sqrt{S_2^2-S_1^2}/r^3$. For a maximally spinning 
hole $2$ and a hole $1$ with a relatively small spin 
at $1000M$ of separation, we obtain a flip-flop period
\be
T_{ff}=\frac{2\pi}{\Omega_{ff}}=32,700\,yr\left(\frac{r}{1000M}\right)^3
\left(\frac{M}{10^8M_\odot}\right),
\ee
which is shorter (by a factor of $40$)
than the gravitational radiation periods reported 
in \cite{Miller:2013gya} used to compare with the accretion-driven 
alignment mechanisms~\footnote{Note that while the
gravitational radiation time scale is $\sim r^4$, the flip-flop time 
scale is $\sim r^3$ (it is a conservative leading term).}.
We thus conclude that such alignment processes might be less effective 
than expected when the flip-flop of spins is taken into account.

These flip-flop configurations might be very effective at
disrupting the inner accretion disk dynamics and at 
circumventing the spin alignment (and growth) process by accretion,
thus leading to important observational consequences.
For instance, the change of the location of the internal rim of the disk 
due to the flip of the spin will change
the high frequencies end of fluctuations and 
the electromagnetic spectrum due to changes
in the efficiency of the conversion of the accreting flow, 
i.e. proportional to $E_{ISCO}(\pm\,a)$. 
Flip-flopping spins might also generate turbulent accretion 
by changing the stirring leading to 
increase/decrease of the radiation (See \cite{Nixon:2013qfa}).
These examples provide rough estimates of the disrupting 
effects of a flip-flopping spin and a more accurate 
evaluation requires a full numerical magnetohydrodynamic 
simulation of such binary black hole configurations.
Our full numerical run proves that, although demanding, 
these simulations are currently possible and they can be 
performed adding a magnetohydrodynamic description of the matter 
on a dynamical binary black hole background~\cite{Noble:2012xz}.

The change in the spin orientation at the latest
stage of the merger could be followed through detailed observation
of the gas jets in X-shaped radio galaxies~\cite{Leahy92}.
The time scale for this phenomena, for instance, for 
the $\sim25000M$ semiperiod we observe for the flip-flop in our full numerical
simulation, corresponds to 1.2 seconds for $10M_\odot$ binaries and
142 days for $10^8M_\odot$ binaries. 
Note that according to Eq.~(\ref{Omegaff})
frequencies can be even higher if the black hole $2$ would be closer
to maximally spinning.


\begin{figure}
\centerline{
\includegraphics[width=\columnwidth]{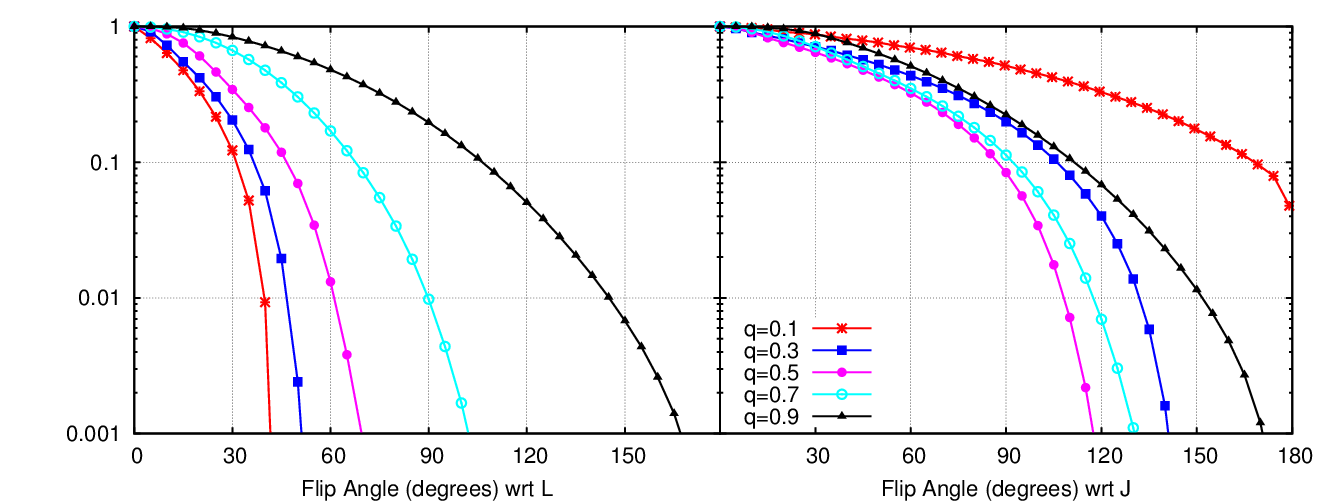}}
\caption{The probability of a spin flip-flop angle $\theta_{ff}\geq x$
for a given mass ratio $q$ and
assuming random spin orientations and magnitudes of the primary
 and secondary black holes.
}\label{probabilities}
\end{figure}

To appreciate the astrophysical relevance of this phenomenon
it is important to determine the likelihood of these flip-flop
angles out of all possible generic binary black hole
merger configurations. We hence consider binaries with
different mass ratios, $q$, and initial random spins
$\alpha_1$, $\cos(\theta_1)$, $\alpha_2$, $\cos(\theta_2)$, 
with $\phi_1-\phi_2=0,\pi$
(this last piece due to the resonances studied in
\cite{Kesden:2010ji,Berti:2012zp,Gerosa:2013laa}).
We evolve these configurations from separations $r=100M$
down to $r=5M$, representing merger, using the 3.5 post-Newtonian approximation.
The results of 2,922,656 simulations per $q$ displaying the probability
of a flip-flop angle larger than $x$ are summarized in Fig.~\ref{probabilities}.
The spin-flip angles remain large for comparable masses and this phenomena
may also occur, to a lesser extent, in black hole - neutron stars
binaries. Note that accretion onto black hole binaries
tends to bring the mass ratio towards 1 because the smaller 
hole is further away from the center of mass of the system 
and can sweep out more mass from the internal parts of the 
circumbinary accretion disk~\cite{Farris:2013uqa}.
The flip-flop frequency for large binary separations $r$ and $q\not=1$
is given by $\Omega_{ff}(q)\approx(3/2)(1-q)/(1+q)(M/r)^{5/2}$.
A thorough study of the unequal mass binary regime is being completed
and will be published by the authors elsewhere.


\begin{acknowledgments}
The authors would like to thank M.Campanelli, J.Krolik, H.Nakano, S.Noble, 
and Y.Zlochower for comments on the original manuscript. Authors also
gratefully acknowledge the NSF for financial support from Grant
PHY-1305730. Computational resources were provided by XSEDE allocation
TG-PHY060027N, and by the BlueSky Cluster 
at Rochester Institute of Technology, which were supported
by NSF grant No. AST-1028087, and PHY-1229173.
\end{acknowledgments}

\bibliographystyle{apsrev4-1}
\bibliography{../../../Bibtex/references}

\end{document}